\documentclass[a4paper,fleqn]{cas-sc}
\usepackage{soul}
\usepackage[sort&compress,numbers]{natbib}
\usepackage{lineno}
\usepackage{setspace}
\usepackage{verbatim}
\usepackage{graphics}
\usepackage{graphicx}
\usepackage{cuted}
\usepackage{lipsum}
\usepackage{multirow}
\usepackage[version=4]{mhchem}
\usepackage{chemformula}

\def\tsc#1{\csdef{#1}{\textsc{\lowercase{#1}}\xspace}}
\tsc{WGM}
\tsc{QE}
\tsc{EP}
\tsc{PMS}
\tsc{BEC}
\tsc{DE}

\begin{document}

\let\WriteBookmarks\relax
\def\floatpagepagefraction{1}
\def\textpagefraction{.001}
\shorttitle{A Concise Review of Recently Synthesized 2D Carbon Allotropes}
\shortauthors{Tromer \textit{et~al}.}

\title [mode = title]{A Concise Review of Recently Synthesized 2D Carbon Allotropes: Amorphous Carbon, Graphynes, Biphenylene and Fullerene Networks}

\author[1]{Ricardo Paupitz}

\author[2]{Alexandre F. Fonseca}

\author[3]{Mizraim Bessa}

\author[2]{Guilherme S. L. Fabris}

\author[4]{William F. da Cunha}

\author[3]{Leonardo D. Machado}

\author[5,6]{Marcelo L. Pereira Junior}

\author[4,7]{Luiz A. Ribeiro Junior}

\author[2,8]{Douglas S. Galvão}

\address[1]{Department of Physics, IGCE, São Paulo State University (UNESP), Rio Claro, São Paulo 13506-900, Brazil.}

\address[2]{Universidade Estadual de Campinas (UNICAMP), Instituto de Física Gleb Wataghin, Departamento de Física Aplicada, 13083-859, Campinas, SP, Brazil.}

\address[3]{Department of Physics, Federal University of Rio Grande do Norte (UFRN), Natal, Rio Grande do Norte 59078-900, Brazil.}

\address[4]{Institute of Physics, University of Brasília (UnB), Brasília, Federal District 70910-900, Brazil.}

\address[5]{Department of Electrical Engineering, College of Technology, University of Brasília (UnB), Brasília, Federal District 70910-900, Brazil.}

\address[6]{Department of Materials Science and NanoEngineering, Rice University, Houston, Texas 77005, United States.}

\address[7]{Computational Materials Laboratory (LCCMat), Institute of Physics, University of Brasília (UnB), Brasília, Federal District 70910-900, Brazil.}

\address[8]{Department of Applied Physics and Center for Computational Engineering \& Sciences, University of Campinas (UNICAMP), Campinas, São Paulo 13083-970, Brazil.}


\begin{abstract}
Two-dimensional (2D) carbon allotropes have received considerable attention due to their unique properties and potential applications in several fields, including electronics, catalysis, energy storage, and sensing. Following the experimental realization of graphene, numerous other 2D carbon structures have been proposed and, in some cases, successfully synthesized. This work presents a concise review of the recently experimentally realized 2D carbon allotropes, including graphynes, biphenylene-based networks, fullerene networks, and monolayer amorphous carbon. For each class, we discuss structural characteristics, theoretical predictions, and synthesis methods, with emphasis on the interplay between theory and experiment. We also highlight instances where experimental studies overlooked relevant theoretical contributions. Finally, we identify theoretically predicted structures that remain unexplored experimentally, suggesting opportunities for synthesis-driven investigations.
\end{abstract}




\maketitle
\doublespacing

\section{Introduction}

Carbon, in addition to being the backbone of organic compounds, stands out for its remarkable ability to form a wide variety of allotropes across multiple dimensionalities and with distinct atomic hybridizations \cite{tiwari2016magical,yadav2025carbon}. These structural forms range from zero-dimensional (0D) fullerenes \cite{mostofizadeh2011synthesis,smalley1997discovering}, one-dimensional (1D) carbon nanotubes \cite{mostofizadeh2011synthesis,shoukat2021carbon} and nanoribbons, two-dimensional (2D) graphene \cite{Novoselov2004} and graphynes \cite{teo_gamma}, to three-dimensional (3D) graphite \cite{bernal1924structure} and diamond \cite{bragg1913structure}, each exhibiting markedly different physical and chemical properties. This versatility makes carbon-based materials highly valuable for a broad range of technological applications, many of which have become the subject of intense recent research.

Among these allotropes, graphene \cite{Novoselov2004} stands out due to its outstanding properties across multiple domains. Despite its simple structure, composed of 2D carbon atoms arranged in a hexagonal lattice, graphene exhibits remarkable electronic and structural characteristics \cite{urade2023graphene}. It is simultaneously lightweight and flexible, while being one of the strongest materials known, enabling its use in mechanically demanding devices \cite{razaq2022review}. Its high electrical \cite{ahmad2023advances} and thermal conductivity \cite{lin2023recent} make it an excellent candidate for nanoelectronic applications, while its optical transparency opens opportunities in photonics \cite{talebzadeh2024review}. Furthermore, its high surface-to-volume ratio renders it attractive for energy storage, catalysis, and even biotechnology \cite{han20223d}.

Due to these attractive properties, graphene has received considerable attention from materials scientists and engineers. Nevertheless, a key challenge limiting its broader applicability in optoelectronics is its lack of an electronic bandgap \cite{wu2023graphene}. While a tunable bandgap is highly desirable for electronic devices, the intrinsic zero-gap nature of graphene restricts its performance as an active layer in devices such as organic light-emitting diodes (OLEDs) \cite{miao2023flexible}, organic photovoltaics (OPVs) \cite{yin2023recent}, and organic field-effect transistors (OFETs) \cite{cai2024comprehensive}. As a result, substantial efforts have been dedicated to discovering alternative materials that could retain many of graphene's desirable properties while exhibiting a suitable semiconducting bandgap for those applications.

In response, both theorists and experimentalists have worked over the years to evaluate, propose, and experimentally realize new 2D carbon allotropes as potential materials for organic electronic devices \cite{jana2021emerging}. Among these structures, graphene nanoribbons (GNRs) \cite{wang2021graphene} stand out. These narrow strips of graphene, obtained by controlled cutting along specific crystallographic directions, exhibit intrinsic properties distinct from those of their parent material. Depending on their symmetry, GNRs can present a non-zero bandgap. However, the synthesis of GNRs remains technically challenging, particularly in controlling the width, length, edge type, and symmetry at a scale and quality compatible with industrial and device-level applications. All these features directly affect the electronic bandgap and transport behavior \cite{houtsma2021atomically}.

Another example is carbon nanotubes (CNTs) \cite{ajayan2001applications}, which can be viewed as rolled-up graphene sheets that form cylindrical structures with periodic boundary conditions along their circumference. While they show promise for nanoelectronic and optoelectronic technologies, the production of high-purity, single-walled CNTs with precise chirality and reproducible structural and electronic properties remains a substantial bottleneck \cite{zhu2022advances}. Thus, beyond possessing a suitable bandgap and preserving the desirable properties of graphene, a third critical requirement emerges: the synthesis process must be reliable, reproducible, and scalable.

In this perspective, given the inherent complexity and evolving nature of synthesis techniques, exploring a broad range of theoretically proposed carbon-based 2D materials has proven to be a fruitful strategy. Indeed, notable progress has been made in recent years by following this path \cite{uddin2023graphene}. This is evidenced by the increasing number of studies focused on 2D carbon allotropes with diverse topologies and atomic arrangements \cite{li2025promising}. Graphene, however, continues to serve as the conceptual and experimental baseline in this area, remaining central to ongoing developments.

This growing interest is well illustrated by the time evolution of scientific publications on the topic. Figure~\ref{fig:fig2} shows the annual number of publications indexed in the Scopus database \cite{scopus} related to graphene and other 2D carbon allotropes. For graphene, a simple Boolean search string was adopted: \texttt{"graphene"}. In contrast, retrieving publications concerning 2D carbon allotropes required a broader bibliographic search strategy to capture their structural and terminological diversity. The following Boolean expression was employed: \texttt{("carbon allotrope" OR "carbon" OR "carbon network" OR "carbon sheet") AND ("2D" OR "two-dimensional" OR "two dimensional" OR "2-dimensional" OR \allowbreak"layered material" OR "monolayer")}.

\begin{figure}[pos=t]
    \centering
    \includegraphics[width=0.5\linewidth]{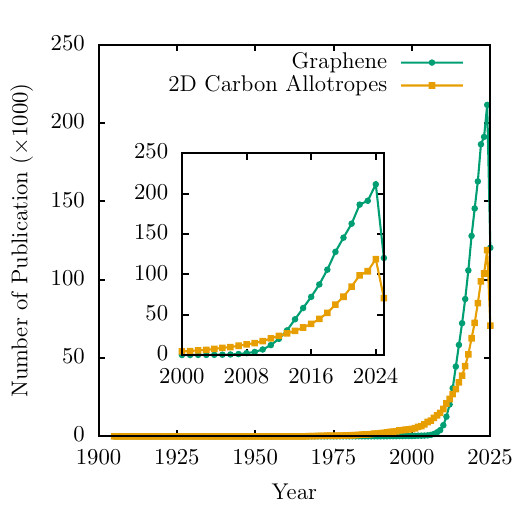}
    \caption{Time evolution of annual publications mentioning graphene and other two-dimensional carbon allotropes, indexed in the \textit{Scopus} database.}
    \label{fig:fig1}
\end{figure}

The data indicate that until 2012, the number of publications related to 2D carbon allotropes surpassed those focusing exclusively on graphene. From 2013 onwards, graphene became the most frequently published topic annually, a trend that remains consistent. As of the latest query, approximately one million documents related to 2D carbon allotropes are indexed in Scopus, while entries explicitly mentioning graphene total around 1.638 million.

This pattern suggests that interest in 2D carbon systems preceded the consolidation of graphene as a central research object. However, the experimental realization of graphene in 2004 \cite{Novoselov2004} marked a significant turning point. It is also important to note that there is considerable overlap between both datasets, as most publications on 2D allotropes cite graphene, often as a theoretical benchmark or reference material. Therefore, the figures should not be interpreted as indicators of competing research areas but rather as evidence of a co-evolutionary and interdependent process. The search for novel carbon allotropes is motivated by the limitations of graphene in specific applications, such as the absence of a bandgap. In contrast, graphene continues to serve as a fundamental material both conceptually and technologically.

In this context, research efforts aimed at developing promising monolayers of 2D carbon allotropes can be generally classified into two major categories. The first category includes studies that report the experimental synthesis of new carbon allotropes without relying explicitly on theoretical predictions. These discoveries are often followed by detailed characterization efforts using both experimental techniques and computational modeling. An example is the case of graphene itself. Although Wallace first described the electronic band structure of a graphene monolayer in 1947 \cite{zhen2018structure}, his work focused on bulk graphite rather than graphene as an isolated 2D system. Later theoretical studies addressed graphene, but often within the scope of specific physical phenomena. For example, Zheng and Ando explored Hall conductivity, referencing graphene only incidentally \cite{zheng2002hall}.

It was not until the experimental realization and fundamental characterization of graphene by Novoselov and colleagues in 2004, \cite{Novoselov2004} that graphene emerged as a central topic in materials research, ultimately leading to the Nobel Prize in Physics in 2010. This milestone triggered a surge of theoretical investigations, including the seminal 2009 review by Castro Neto and collaborators \cite{geim2010nobel}, which thoroughly examined the electronic properties of graphene.

The second research pathway is more result-oriented and involves experimental synthesis efforts that are explicitly guided by theoretical predictions. This strategy reduces the exploratory cost by targeting materials that have already been identified as promising through simulations. A notable example is the experimental synthesis of graphene nanoribbons, driven by the theoretical insights of Mitsutaka Fujita in 1996 \cite{fujita1996peculiar}, who predicted distinctive electronic behavior arising from edge topologies.

\begin{figure}[pos=htb]
    \centering
    \includegraphics[width=0.7\linewidth]{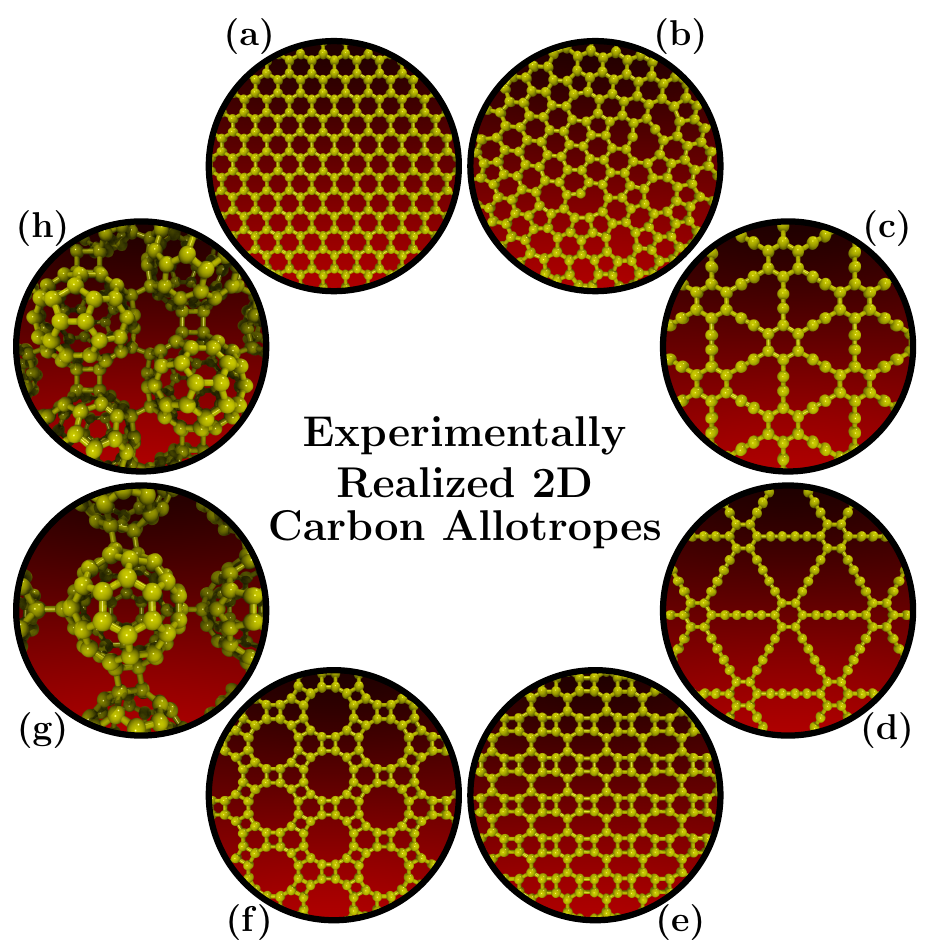}
    \caption{Experimentally realized 2D carbon allotropes. The panels represent different structures:
    (a) graphene, 
    (b) monolayer amorphous carbon (MAC),
    (c) $\gamma$-graphyne,
    (d) graphdiyne,
    (e) biphenylene network,
    (f) graphenylene network,
    (g) quasi-hexagonal phase C$_{60}$ (qHPC$_{60}$),
    (h) quasi-triangular phase C$_{60}$ (qTPC$_{60}$).}
    \label{fig:fig2}
\end{figure}

It is essential to note that while studies employing the second approach generally acknowledge prior experimental work that validates their predictions, the reverse is not always the case. Experimental studies sometimes fail to cite earlier theoretical contributions that were available at the time of their discovery. This gap often reflects a lack of integration between theoretical and experimental communities. Closing this gap could accelerate progress in the field.

In this work, we present a critical review of experimentally realized 2D carbon allotropes, emphasizing their atomic structures, synthesis routes, and physical properties. While the primary focus lies on experimentally realized systems, we also consider theoretical contributions that have either guided their discovery or proposed related nanostructures with potential technological relevance. The review is organized as follows. The next section examines monolayer amorphous carbon (MAC). A discussion on graphyne and graphdiyne lattices follows this. Subsequently, biphenylene-based carbon networks (BPN) are analyzed, and 2D fullerene-derived structures are presented. The manuscript concludes with final remarks and perspectives on current challenges and future directions for the field. An illustrative overview of the experimentally realized 2D carbon allotropes covered in this review is presented in Figure~\ref{fig:fig2}. The figure highlights the structural diversity of these materials, including graphene, MAC, graphyne derivatives, biphenylene-based networks, and fullerene monolayers.

\section{Pristine and Amorphous\\Monolayer Carbon}
\label{section:mac}

MAC is one of the most notable examples among the recently synthesized 2D carbon allotropes. It consists of a single atomic layer of carbon atoms arranged in a disordered fashion, lacking long-range crystalline order. Despite its amorphous nature, MAC shares several key properties with graphene, including high electrical conductivity, optical transparency, large surface area, and mechanical strength. Importantly, it exhibits a small electronic bandgap, which distinguishes it from pristine graphene. These features have attracted growing interest from both experimental~\cite{Toh2020SynthesisAP, PhysRevLett.106.105505} and theoretical~\cite{https://doi.org/10.1002/pssb.200945581, PhysRevB.84.205414,hu2025simulated,dos2025exploring,felix2020mechanical,junior2021reactive,tromer2021optoelectronic,tromer2024structural} communities.

Unlike crystalline solids, amorphous materials do not exhibit long-range periodicity in their atomic arrangement~\cite{drabold2009topics,wilding2006structural}. Although localized regions may resemble crystalline motifs, finite unit cells are insufficient for capturing their structure fully. Techniques such as radial distribution function (RDF) analysis are therefore widely used. Amorphous materials have found broad application in diverse scientific and technological contexts~\cite{ovshinsky1987progress,meydan1994application,zhou2020thermal,li2019amorphous,wang2025turning}.

A foundational model for such systems was introduced in 1932 by Zachariasen~\cite{zachariasen1932atomic}, who proposed the continuous random network (CRN) framework. This model describes disordered yet connected atomic arrangements lacking translational symmetry, and it remains central to understanding amorphous structures~\cite{https://doi.org/10.1002/adma.201301909}. Nevertheless, recent studies suggest deviations from the CRN model in specific systems, motivating the development of refined approaches.

A breakthrough in the field occurred in 2019 when Toh \textit{et al.} synthesized and characterized MAC~\cite{Toh2020SynthesisAP}. Using laser-assisted chemical vapor deposition (CVD), a continuous monolayer was formed within minutes at relatively low substrate temperatures (approximately 200~$^\circ$C), in contrast to the higher temperatures typically required for graphene growth~\cite{joo2017realization}.

High-resolution transmission electron microscopy (HRTEM) confirmed the absence of crystalline periodicity, revealing a disordered carbon network. The structure is best described by a crystallite-in-Z-CRN model, where nanoscale domains of six-membered rings are embedded in a continuous random network. These domains vary in orientation and strain. The atomic arrangement includes pentagons, heptagons, octagons, distorted hexagons, and regular hexagons, the latter forming crystalline inclusions of approximately 1 nm in diameter. Bond lengths ranged from 0.9 to 1.8~\r{A}, and bond angles spanned 90$^\circ$ to 150$^\circ$, reflecting significant structural variability.

\begin{figure}[pos=htb]
    \centering
    \includegraphics[width=0.3\linewidth]{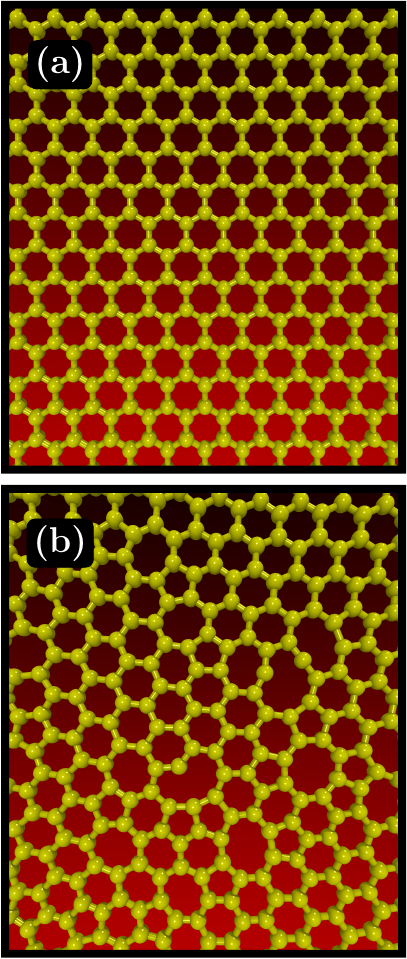}
    \caption{Atomistic representations of (a) pristine graphene and (b) monolayer amorphous carbon, illustrating the contrast between the long-range crystalline order of graphene and the disordered topology of MAC.}
    \label{fig_GR_MAC}
\end{figure}

The material also exhibited excellent film continuity, with no multilayer regions or wrinkles, and remarkable environmental stability, maintaining its structure for over a year. This contrast between the ordered lattice of graphene and the MAC disordered topology is illustrated in Figure~\ref{fig_GR_MAC}(b).

Theoretical modeling of MAC was performed using a kinetic Monte Carlo approach on a $40 \times 40$~\r{A}$^2$ supercell with 610 carbon atoms, followed by conjugate gradient relaxation. The AIREBO potential~\cite{stuart_JCP} was used for interatomic interactions, implemented in LAMMPS~\cite{plimpton_JCP}. Among over 60,000 configurations, the one that best matched the HRTEM features was selected as the reference structure.

This model has been employed in multiple studies to explore MAC's mechanical, electronic, optical, and thermal properties~\cite{felix2020mechanical,tromer2021optoelectronic,pereira2022mechanical,xie2021roughening,xie2023toughening}. Complementary density functional theory (DFT) calculations using VASP~\cite{kresse1996efficient} and transport simulations with TranSIESTA~\cite{brandbyge2002density,soler2002siesta} confirmed its room-temperature thermal stability and revealed buckling in non-hexagonal regions. An optical bandgap of approximately 2.1~eV was also identified, underscoring MAC's promise in optoelectronics.

Following this foundational characterization, we extended the analysis to assess MAC's mechanical and thermal behavior in comparison to graphene~\cite{felix2020mechanical}. Simulations using the same 610-atom MAC model and a 640-atom graphene structure were subjected to uniaxial tension up to a strain of 100\%. MAC exhibited a Young's modulus ($Y_M$) of approximately 323 GPa, compared to 881 GPa for graphene, under the same conditions. While graphene reached a tensile strength of approximately 228 GPa, MAC failed at around 61 GPa. The disordered MAC structure allows more bond rearrangement and accommodates strain more gradually, leading to extensive linear atomic chain (LAC) formation upon rupture.

Thermal ramp simulations, spanning from 100 K to 10,000 K, revealed collapse mechanisms in both systems. LAC formation was observed at temperatures of approximately 4500~K for MAC and 5500~K for graphene. Melting points were estimated to be approximately 3600 K for MAC and 4100 K for graphene.

To explore curvature effects, further investigations examined MAC configured as nanotubes and nanoscrolls~\cite{tromer2021optoelectronic}. Density Functional-based Tight Binding (DFTB) and DFT calculations revealed that rolling the sheet resulted in small bandgap openings (90, 13, and 18 meV for tubes and scrolls). These systems absorb light from the infrared (IF) to ultraviolet (UV), with reflectivity decreasing as the wavelength moves toward the UV. Scrolls showed the largest optical response.

Using larger supercells (3660 atoms), molecular dynamics simulations assessed the thermal and mechanical properties of MAC-based nanotubes and nanoscrolls~\cite{pereira2022mechanical}. Results confirmed that MAC exhibits comparable mechanical behavior to its crystalline counterpart under curvature. Melting temperatures varied between 5100 K and 5900 K, depending on the geometry and strain distribution.

\section{Graphyne and Graphdiyne Allotropes}
\label{section:graphynes}

One of the first predictions of a one-atom-thick 2D carbon allotrope that is distinct from graphene is the structure named graphyne (GY). 
In 1987, Baughman, Eckhardt, and Kertesz~\cite{Baughman19876687} proposed seven new structures of periodic 2D carbon phases formed by acetylenic chains (\ch{\bond{single}C+C\bond{single}}), connecting aromatic rings or carbon-carbon bonds. 
These can be conceptualized as systematic substitutions of carbon-carbon bonds in graphene by acetylenic groups. 
There are innumerable ways to generate structural GY motifs by connecting acetylenic chains directly to themselves or through aromatic rings and sp$^2$ carbon-carbon bonds. 
Among the seven GY structures originally proposed, those designated as $\gamma$-, $\beta$-, and $\alpha$- and 6,6,12-GYs are the most investigated.
The first three members of the graphyne family exhibit hexagonal symmetry, with $\gamma$-GY being the densest of the graphyne family. 

\begin{figure}[pos=t]
    \centering
    \includegraphics[width=0.3\linewidth]{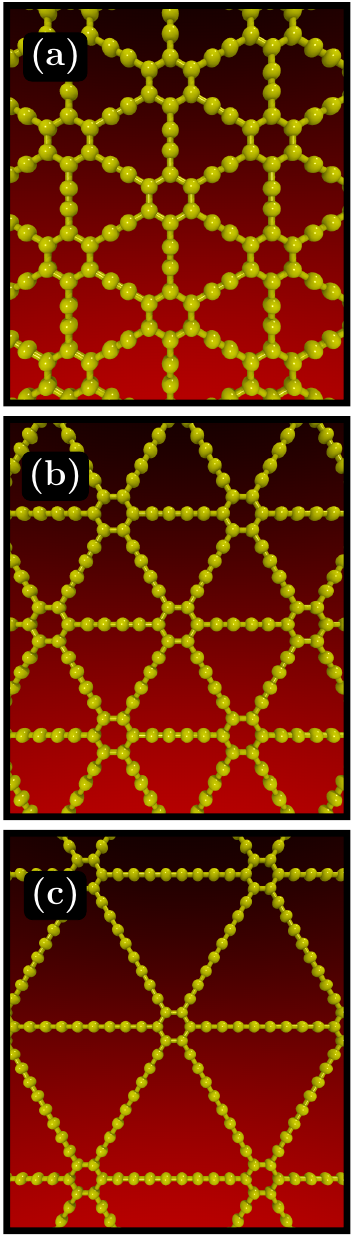}
    \caption{Top views of $\gamma$-graphyne $(n=1)$, $\gamma$-graphdiyne $(n=2)$ and $\gamma$-graphtetrayne $(n=4)$. }
    \label{fig_GY_GDY}
\end{figure}

The GY structures can be created with different numbers of acetylenic chains, $n$ (\ch{\bond{single}C+C\bond{single}}$_n$, $n\geq1$), without changing the structural symmetry. 
Structures having $n=2, 3, 4,$ are designated as {graphdiyne} (GDY), {graphtriyne}, {graphtetrayne} (GTY), and so forth, respectively. 
Figure~\ref{fig_GY_GDY} illustrates the structures of $\gamma$-GY, $\gamma$-GDY and $\gamma$-GTY structures. 

\subsection{Theoretically Predicted Properties}
\label{theoGY}

Interest in GYs emerged decades later, based on several works that predicted that the physical properties of GYs are not only as special as those of graphene~\cite{jun2018REVIEW-ACSAMI}, but maybe even more interesting. 
For instance, similar to graphene, it has been demonstrated that GYs can exhibit high charge carrier mobility~\cite{ChenJPCLett2013}, relatively good thermal conductivity~\cite{MingJCP2013}, and strong mechanical properties~\cite{buhelerCARBON2011,ZhangAPL2012}. 
Similar to graphene, GYs possess several unique properties. 
They are porous~\cite{kanegaeMRS2023}, exhibit a non-zero electronic bandgap~\cite{Narita1998,MalkoPRL2012,WuNANOSCALE2013}, which is an important feature for electronic applications. 
In addition, their electronic properties can be tuned~\cite{PadilhaJPCC2014}, which allows the design of new applications in electronic devices. Finally, asymmetric GYs display null and negative linear compressibilities~\cite{kanegaeCARBONTRENDS2022,kanegaeMRS2025}, a property that could be exploited for applications such as sensors and other mechanical actuators. 
These properties underscore the rationale for considering GYs as a substitute for graphene in electronic and other applications~\cite{SchirberPHYSICS2012,JonSCIENCE2012}.

Recent literature has predicted that GYs possess additional attractive properties, including negative thermal expansion~\cite{Cheol2015PRB,Hernandez2017}, Poisson's ratio greater than 0.5~\cite{Hernandez2017,kanegaeCARBONTRENDS2022}, interesting optoelectronic properties~\cite{DaniMSSP2025}, magnetic hystereses~\cite{KantarPHYSSCRIPTA2024}, thermoelectric features~\cite{WangJCP2014,SevickNANOTECH2014}, enhanced elastocaloric effect~\cite{kanegaeACSAMI2025} and thermal stability up to about 1000 K~\cite{SolisACSAMI2019}. 
A number of computational studies have been conducted to explore the existence and properties of GY scrolls~\cite{SolisACSAMI2019}, nanoribbons~\cite{LiuPCCP2023,kanegaeACSAMI2025}, and nanotubes~\cite{ColuciPRB2003}. 

These theoretical works have contributed to the study of potentially new applications of GYs. 
Examples include the development of nanotransistors~\cite{LongACSNano2011}, gas separation and desalination~\cite{LinNANOSCALE2013,KouNANOSCALE2014,YeoADVMATER2019,QiuADVMATER2019,AziziSCIREP2021}, anticancer and antibacterial agents~\cite{DebASS2023,WangNM2025}, anodes for batteries~\cite{SakamotoADVMATER2019,BartolomeiML2024}, hydrogen storage~\cite{QuejCMS2024}, supercapacitor electrodes~\cite{ChenJPHYSCM2020,HenriqueENESTOR2025}, and others~\cite{Ivanovskii2013,LiCHEMSOCREV2014,LuanJPCB2016,LiGIANT2023}.

\subsection{Synthesized GYs}
\label{synGY}

Following the seminal contributions of Baughman, Eckhardt, and Kertesz~\cite{Baughman19876687}, numerous attempts have been documented in the literature to synthesize GYs. 
The synthesis of GY oligomers and ethynylene-linked molecules has been achieved. However, the yield has been limited to small fragments~\cite{CallstromMACRO1988,EislerANGEW1999,kehoeOL2000,ZhangJOC2005,JohnsonOL2007,HaleyPAC2008,DiederichADVMAT2010}.

The first successful synthesis of GYs was reported in 2010 by Liu \textit{et al.}~\cite{LiuCHEMCOMM2010}, for $n=2$ or the $\gamma$-GDY. 
The synthesis of GDYs on a Cu foil was accomplished through the implementation of an \emph{in situ} Glaser-Hay coupling reaction~\cite{Glaser1,Glaser2,Hay} of hexaethynylbenzene (HEB) monomers with pyridine as the organic solvent. 
The efficacy of Liu's methodology for obtaining GDYs can be attributed to the catalytic properties of Cu ions dispersed in pyridine, which facilitate the Glaser reaction of the HEB monomers. Additionally, the Cu foil serves as a template for GDY growth, contributing to the method's success. 
%

After the initial report by Liu {\it et al.}, research on GDYs has undergone exponential growth, encompassing the development of novel growth methods~\cite{matsuoka2017JACS,Zuo2017ChemComm,Zhou2018ACSAMI,GaoCHEMSOCREV2019,Yin2020AdvFuncMat}, with a remark on a recent report on the achievement of ultra-fast, second long, synthesis of large amounts of GDY from electron beam irradiation on HEB--hexakis-[(trimethylsilyl)ethynyl]benzene compounds under copper(II) acetate and N,N-dimethylformamide as the catalyst and solvent, respectively~\cite{Kuang2025AdvMat}.

The synthesis of graphtetrayne (GTY), i.e., a GY with $n=4$, was reported almost a decade later by Gao \textit{et al.} ~\cite{GaoNANOENE2018} and Pan  \textit{et al.} ~\cite{PanCCSCHEM2021}. These works have received comparatively less attention in the literature than those of GDYs, which may be attributable to the challenges associated with their synthesis.   
The Sonogashira reaction method~\cite{RafaelCHEMREV2007} was employed in both works to describe an intermediate step analogous to that utilized in the synthesis of GDYs. 
This step involves the synthesis of monomers, which are composed of a carbon ring with chains of four acetylenic arms. 
Subsequently, the GTYs were grown on copper foils through reactions analogous to Glaser-Hay.

The synthesis of GDYs and GTYs represents a significant breakthrough in the field. 
However, their large porous nature imposes certain limitations on their properties. 
Nevertheless, the difficulties associated with synthesizing graphyne with $n=1$, have been recently surmounted. 
Li  \textit{et al.} ~\cite{LiCARBON2018} were the first to report the synthesis of GYs with $n=1$, through a mechanochemical synthesis method.

Four years later, in 2022, Barua, Saraswat, and Rao~\cite{BaruaCARBON2022}, and Rodionov {\it et al}.~\cite{RodionovJACS2022}  reported successful synthetic approaches, based on Sonogashira crosslinking reactions, for the production of the densest form of GY, the $\gamma$-GY. 
The synthesis of another form of GY with $n=1$, termed Holey $\gamma$-GY, has also been reported by Liu et al. ~\cite{LiuMATTER2022}. 
In a recent study, Song {\it et al.}~\cite{SongACIE2024} reported the wet chemistry synthesis of $\gamma$-GY based on a metal-free crosslinking method. This method involves the combination of fluoro-(hetero)arenes and alkynyl silanes in the presence of tetrabutylammonium fluoride. 
According to Song {\it et al.}, their method is capable of producing $\gamma$-GY, ranging from grams to the sub-microscopic level.

Li and Han~\cite{LiGIANT2023} posit that among the various methods of synthesizing GYs, the mechanochemical approach is suboptimal due to its inability to adequately control reactions and remove impurities, resulting in substandard product quality.

In a recent study, Rodionov {\it et al.}~\cite{RodionovCARBON2025} tried to replicate the mechanochemical method for synthesizing GYs. 
The validity of the achievement was contested by demonstrating that the application of the identical method under equivalent conditions yielded disordered structures without the presence of triple bonds between carbon atoms. 
Consequently, the Sonogashira reaction methods can be regarded as the most suitable approach to obtain high-quality $\gamma$-GY samples~\cite{LiGIANT2023}. 
Nevertheless, Li and Han~\cite{LiGIANT2023} have noted that a methodology for producing GYs with a variable number of layers, size, and stacking modes remains to be developed.

The synthesis of the other members of the GY family remains to be achieved.  However, the synthesis of $\gamma$-GY could potentially expedite the identification of a method to produce the other structures. 
The synthesis of the Holey $\gamma$-GY~\cite{LiuMATTER2022} and the recent production of precursors of the 6,6,12-GY~\cite{KildeNATCOMM2019,KrollCEJ2025} suggest that the synthesis of more $n=1$ GYs would be possible.  

In relation to the thermal metastability of GYs with $n=1$~\cite{SolisACSAMI2019}, a recent study has demonstrated that stacks of $\gamma$-GYs undergo an irreversible transformation into planar, nongraphitic, pure sp$^2$ carbon crystals~\cite{RayPNAS2025}. 
The transformation occurs exothermically at ambient pressure, with a temperature range of 160  $^\circ$C to 310 $^\circ$C. 
This finding suggests a novel approach for synthesizing new 2D carbon phases.

\section{Biphenylene-Based\\Carbon Networks}
\label{section:biphenylene-based}

The molecule known as biphenylene is formed by two hexagonal carbon rings connected by two covalent bonds that define a 4-membered ring. Due to its single and double bonds configuration, it combines aromatic and anti-aromatic properties\cite{Totani2017,vogtle1992fascinating}, being at the same time chemically stable\cite{holm2011dissociation} and presenting interesting physicochemical properties, which several authors explored in the last decades\cite{elsaesser1988picosecond,Mak1962,beck1997vertical,zimmermann1996theoretical,Liu2021}. This, along with the rise of interest in 2D materials, motivated several groups, both theoretical and experimental, to investigate the possibility of obtaining technologically promising materials based on periodically arranged 2D structures formed using biphenylene as the repeating unit. It is worth noting that, in 1968 and 1989, Balaban and collaborators proposed several carbon-based structures that were later ``reinvented'' with different names in more recent studies. Among the various theoretically predicted configurations, one is the BPN \cite{balaban1968chemical,balaban1989carbon}.

\subsection{Theoretical Predictions for \\ Biphenylene-based 2D Materials}

\begin{figure}[pos=t]
    \centering
    \includegraphics[width=0.3\linewidth]{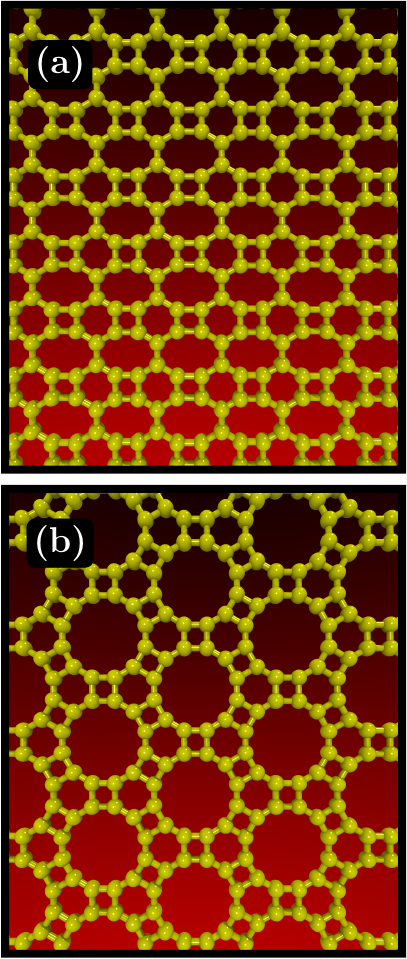}
    \caption{Two experimentally obtained biphenylene-based 2D materials. (a) Biphenylene Network, where it is possible to identify the four-, six-, and eight-membered rings. (b) Graphenylene lattice, composed of four-, six-, and twelve-membered rings.}
    \label{fig:BPN_BPC}
\end{figure}

Different 2D lattices can be constructed by connecting biphenylenes; two of them have already been synthesized and are exemplified in Figure \ref{fig:BPN_BPC}. The material known as BPN is shown in figure \ref{fig:BPN_BPC}(a), where we can see it is composed of four-, six-, and eight-membered rings forming a rectangular lattice in which the repeating unit is a biphenylene molecule. The lattice constants are 3.75 \r{A} and 4.52 \r{A} in the horizontal and vertical directions in figure \ref{fig:BPN_BPC}(a), respectively. The lattice is found to be relatively stable, with cohesive energy E$_\text{coh}=-7.55$ eV/atom, although being less stable than graphene, and its work function is estimated as $\Phi=4.30$ eV\cite{BPN_first_principles-2021}.
BPN is predicted to be a better fit for lithium-ion batteries than most 2D carbon materials, since the adsorption of that metal would be more energetically favored than in graphene or graphyne, for instance \cite{biphenylene_H_storage_2015}.

As an experimentally synthesized material, the interest in BPN-like structures has increased.  In the last few years, several studies have been reported on experimental methods for their synthesis and possible technological applications \cite{biphenylene_exp_2014,biphenylene_synth_uses_2019,biphenylene_network_2021,biphenylene_synthesis_2025,biphenylene_review_2025}.

It is interesting to note that these materials behave like a metal when presented as a sheet or a nanotube. In contrast, they become semiconductors when functionalized or cut as nanoribbons\cite{biphenylene_ribbons_2010,Biphenylene_ribbons_functional_2014}. This tunability of physical properties makes them versatile candidates for various applications, including gas storage, gas separation, and optical sensors\cite{biphenylene_excitonic_2016}, as well as nanoelectronic devices and nanosensors of different types.

On the theoretical side, studies regarding BPN have shown that armchair and zigzag nanoribbons \cite{BPN-optical-catalytic-2024} have distinctive electronic and optical properties compared to other carbon allotropes.
Armchair nanoribbons can present conducting or semiconducting character, with high mobility ($\sim57.174~\text{cm}^2\,\text{V}^{-1}\,\text{s}^{-1}$) depending on the number of unit cells
\cite{BPN-nanoribbon-armchair-2016}.
GW and Bethe-Salpeter-based calculations suggest that, for BPN nanoribbons,  absorption peaks due to excitons are located in the visible range. Additionally, the more relevant bound excitons are found to present a strong electron-hole interaction, which can be relevant for photovoltaic applications\cite{BPN-Optical-2016,BPN-optical-catalytic-2024}.

Other works suggest their electronic properties can be engineered, achieving a strain-induced metal-insulator transition at relatively high strain values, opening a $\Gamma$-X indirect electronic band gap\cite{Biphenylene-network-thermoelectric}. 
Furthermore, BPN networks demonstrate high mechanical stability and flexibility \cite{BPN-fracture-2022,RAHAMAN201765}. Also, the high anisotropy of its thermoelectric properties has been highlighted \ cite {Biphenylene-network-thermoelectric}, reinforcing its position as a promising material for electronic applications. 

Another relevant 2D lattice created using biphenylenes and already experimentally realized is shown in figure \ref{fig:BPN_BPC}(b), known as Graphenylene (GP). This material features a unique geometrical structure composed of four-, six-, and twelve-membered rings. Its unit cell belongs to the P6/mmm space group with one irreducible atom. The central nanopore, formed by a dodecagon ring, has a diameter of 5.47  \r{A}\cite{Fabris2018}. Its cohesive energy ($-9.7$ eV) is close to values found for graphene ($-10.4$ eV), suggesting slightly lower  stability\cite{FABRIS2021}.

The GP structure reveals seven points of high symmetry, including hollow adsorption sites within the rings and edges between different rings. These sites are critical for applications such as gas separation and sensing, as they influence the diffusion barriers for ions, atoms, or molecules passing through pores. The interaction of chemical species with these sites can also influence the adsorption energies of small molecules or atoms interacting with the membrane, thereby altering the material's adsorption capacity. Mechanically, GP demonstrates remarkable properties, surpassing some 2D materials due to its symmetry-driven elastic constants, which yield a $Y_M$ of 648.8 GPa and a Poisson's ratio ($\nu$) of 0.259. While GP's modulus is lower than graphene's (1 TPa), its mechanical versatility remains significant\cite{Fabris2018}. In the literature, GP is reported to present a direct electronic bandgap below  1eV\cite{Brunetto1212810,graphenylene2013} at the K point, differing fundamentally from graphene's Dirac cone structure. Its electronic bands, being almost parabolic close to the K point, result in excellent electronic mobility, with effective masses for electrons (m$^{*}$/m$_{e}$ = 0.26) and holes (m$^{*}$/m$_{h}$ = 0.33), making GP a promising intrinsic semiconductor\cite{Brunetto1212810}.

This 2D porous material exhibits two important characteristics \cite{Brunetto1212810} that should be observed in an ideal carbon-based 2D semiconductor: on one side, an intrinsic electronic bandgap close to that of usual semiconductors; on the other side, GP presents a reasonable electronic dispersion, resulting in electronic delocalization and high charge mobility.
Excitonic properties of GP were also investigated, revealing that its optical excitation can lead to bright and dark excitons with binding energies of 530 meV, indicating BP as a candidate for applications in solar energy and quantum information technology\cite{exciton-graphenylene2016}.

The dodecahedral pore of GP exceeds the kinetic diameter of certain molecules, enabling efficient gas diffusion. Song \textit{et al.} \cite{Song2013} concluded, via DFT calculations, that H$_{2}$ exhibits a low energy barrier (0.20 eV) for diffusion, unlike CO (0.99 eV), CO$_{2}$ (1.05 eV), and N$_{2}$ (1.01 eV). GP also shows promise in cryogenics, particularly for separating helium isotopes ($^{3}$He and $^{4}$He) due to its high permeance and quantum tunneling properties. It was also confirmed that GP outperforms inorganic graphenylene (IGP) \cite{Perim2014}, with superior separation factors and permeance ($10^{-8}$ mol\,m$^{-2}$\,s$^{-1}$\,bar$^{-1}$) at high temperatures\cite{Motallebipour2021}. Moreover, improved separation efficiency under steady-state conditions poses IGP as more effective than GP at elevated temperatures. Strain tuning further enhances GP's performance. Zhu \textit{et al.}\cite{Zhu2016} demonstrated a sixfold increase in H$_{2}$ permeance ($2.6 \times 10^{-2} ~\text{mol}\,\text{s}^{-1}\,\text{m}^{-2}\,\text{Pa}^{-1}$) under 3.04\% strain, strains of 3.04\%-4.20\% enable CO$_2$ separation, while $5.12$-$10.78\%$ improves methane (CH$_4$) separation. These findings suggest GP has great potential as a tunable material for technological applications.

Addressing high non-renewable energy consumption and toxic gas removal is critical, and carbon-based catalysts, enhanced by anchored metals, are effective in mitigating atmospheric pollutants.\cite{Liu2016,PrezMayoral2016} Chen \textit{et al.}\cite{Chen2020} demonstrated via first-principles calculations that GP doped with single Ru and Mo atoms, facilitates CO and NO oxidation through the Eley-Rideal (ER) mechanism, which is more efficient than the Langmuir-Hinshelwood (LH) model. Tang \textit{et al.}\cite{Tang2021} showed that Pt and Pd anchored on GP exhibit low energy barriers ($<0.35$ eV) for the ER mechanism during CO$_{2}$ and N$_{2}$ production. Transition metals (Mn, Co, Ni, Cu) also function effectively as single-atom catalysts on GP surfaces\cite{Tang2020}. Furthermore, Fe anchored on GP demonstrates strong catalytic performance in CO oxidation and multifunctional gas sensing (CO, NO, NO$_{2}$, O$_{2}$) due to its magnetic and electronic properties.

The limitations of traditional anode materials, such as graphite, with a specific capacity of $372 \, \text{mAh} \, \text{g}^{-1}$, highlight the need for alternatives. Yu\cite{Yu2013} demonstrated that graphenylene (GP) offers superior lithium mobility and storage capacities --- $1116 \, \text{mAh} \, \text{g}^{-1}$ (Li$_3$C$_6$) and $930 \, \text{mAh} \, \text{g}^{-1}$ (Li$_{2.5}$C$_6$) --- through DFT studies. However, single-sided Li adsorption reduces capacity to $487 \, \text{mAh} \, \text{g}^{-1}$. GP also shows promise for sodium-ion batteries (SIBs), achieving $729.89 \, \text{mAh} \, \text{g}^{-1}$ (NaC$_3$),\cite{daSilvaLopesFabris2021} which surpasses graphite ($100 \, \text{mAh} \, \text{g}^{-1}$) and doped graphene ($308 \, \text{mAh} \, \text{g}^{-1}$).\cite{Wasalathilake2018} Additionally, Hussain \textit{et al.}\cite{Hussain2017} found that GP doped with alkali metals stabilizes 20 H$_2$ molecules, enhancing hydrogen storage. In water desalination, fluorinated GP membranes achieve high water permeability ($11.032 \, \text{L} \, \text{m}^{-2} \, \text{h}^{-1} \, \text{bar}^{-1}$) and $99.4\%$ salt rejection, proving their efficacy for water cleaning applications\cite{Jahangirzadeh2022}. Jahangirzadeh \textit{et al.}\cite{Jahangirzadeh2024} showed that GP has an excellent ability to separate salt ions under high pressures, with a permeability of $251.351 \, \text{L} \, \text{m}^{-2} \, \text{h}^{-1} \, \text{bar}^{-1}$ and salt rejection of $98.8\%$.

Another realm in which GP can be considered a promising alternative to standard materials is the construction of nanoscale electronic devices. Villegas-Lelovsky and Paupitz\cite{VillegasLelovsky2020} demonstrated that doping GP enables electronic transport tuning, forming {\sl pn} junctions suitable for low-dimensional devices such as diodes, with a conducting threshold of $\sim1.5 \, \text{eV}$. The possibility of Zener-like configurations was also discussed in that paper. Additionally, \textit{ab initio} calculations highlight GP/MoX$_2$ van der Waals heterostructures (X = S, Te, Se) for diodes with high rectification factors ($10^3$-$10^4$) and optical absorption spectrum in the range of visible light, making them promising for photonics and nanoelectronics.\cite{Meftakhutdinov2021} Liu \textit{et al.}\cite{Liu2015} showed GP's functionalization via hydrogenation and halogenation, achieving band gap energy ranges of 0.075-4.98  eV and 0.024-4.87 eV, respectively, posing GP as a versatile material for band gap engineering.

GP and similar materials have also been identified as high-order topological insulators \cite{Yang2025}. Due to a zero dipolar moment, their characterization is somewhat more sophisticated than that of regular topological insulators, bringing the necessity of taking into account higher order moments, like quadrupoles and octupoles\cite{multipole-insulators-science-2017,Benalcazar2019,Takahashi2021,Yang2025}. Furthermore, these materials are promising for the study of higher-order topological superconductors. Such a rich set of topological properties suggests that this class of materials should be considered in future investigations regarding the proposition of nanodevices, especially those designed for the use of topological properties in quantum computing\cite{Wong2023,Chu2023}.  

\subsection{Synthesis of Biphenylene-based\\2D Materials }

A bottom-up synthesis of the Biphenylene network was attempted for a few years by different groups, resulting in a number of papers that describe interesting but only partial successes.
Despite the existence of theoretical propositions for its practical synthesis, as the use of uniaxial tension to convert penta-graphene into BPN \cite{RAHAMAN201765}, these propositions are yet to be tested in the real world. Over the years, a few works have reported the synthesis of linear structures, or low-width nanoribbons, which exhibit semiconducting behavior, quite different from the metallic characteristics expected for the 2D material\cite{biphenylene_exp_2014,Biphenylene_ribbons_functional_2014}.   

A significant advance in the field was achieved in 2021, when Fan and coworkers\cite{biphenylene_network_2021} successfully utilized an interpolymer dehydrofluorination (HF-zipping) reaction to obtain the material deposited on a gold surface.
Additionally, recent reports have suggested new routes for synthesizing large sheets, such as polymerization at room temperature triggered by pressure application, which can induce further structural transitions at high pressures, namely 3GPa and 14GPa\cite{biphenylene_synthesis_2025}.
Since this field is under development, many possibilities are being investigated and tested both theoretically and experimentally.

Regarding the synthesis of graphenylene, byphenylene-based structures were proposed in 1968, by Balaban and collaborators \cite{Balaban1968}. Its synthesis was later discussed by Baughmann et collaborators, who argued that producing it from graphyne would be thermodynamically unfavorable \cite{Baughman19876687}. Many years later, after the realization of graphene \cite{Novoselov2004}, some groups began to reexamine this structure \ cite {Brunetto1212810,graphenylene2013,koch2015graphenylene}.   

The experimental realization of another porous 2D carbon-based material, achieved in 2009, further fueled interest in the search for new 2D carbon allotropes. This was the synthesis of a new  2D structure with periodically distributed nanopores similar to graphene, one-atom thick, fabricated by Bieri and collaborators\cite{bieri2009porous}. This insulating molecular network was obtained using the hexaiodo-substituted macrocycle cyclohexa-m-phenylene (CHP), along with an aryl-aryl coupling reaction on an Ag(111) substrate. The surface-assisted coupling reaction started after annealing the substrate for a few minutes at 575 K, as indicated by the STM analysis, and the polymerization reaction started above this temperature, resulting in the formation of a completely developed polyphenylene network, where the iodo substituents are eliminated, resulting in the formation of covalent bonds between the macrocycles, leading to an ordered network. Measurements of high-resolution STM of this synthesized polyphenylene superlattice showed a uniform nanopore spacing of 7 \r{A}, with an average pore diameter of 5.8 \r{A}. Interestingly, further theoretical works have suggested a possible transition of the polyphenylene network to a new carbon allotrope, which combines hexagonal and square carbon rings,\cite{Brunetto1212810, FABRIS2017} also known as graphenylene.

Years later, in 2017, Du \textit{et al.}\cite{du1740796} used another chemical route to synthesize this 2D carbon crystal, which had been predicted theoretically, from 1,3,5-trihydroxybenzene. This material, consisting of 4- and 6-carbon atom rings, was named by the authors as 4-6-carbophene. The chemical reactions to synthesize graphenylene involve a polymerization process that can occur through two possible pathways. The first is an intramolecular dehydration of 1,3,5-trihydroxybenzene, where water molecules are removed using aluminum oxide, causing the benzyne intermediates to combine and form small graphenylene fragments. The second possible reaction is an intermolecular dehydration between 1,3,5-trihydroxybenzene molecules, where they combine to form a product through a rapid reaction. This synthesis was performed in an argon atmosphere at 350-380 $^\circ$C in a quartz glass tube furnace using $\gamma$-Al$_2$O$_3$ as the dehydrant agent.

\section{Two-Dimensional\\Fullerene Networks}
\label{section:fullerene-network}

\begin{figure}[pos=t]
    \centering
    \includegraphics[width=0.3\linewidth]{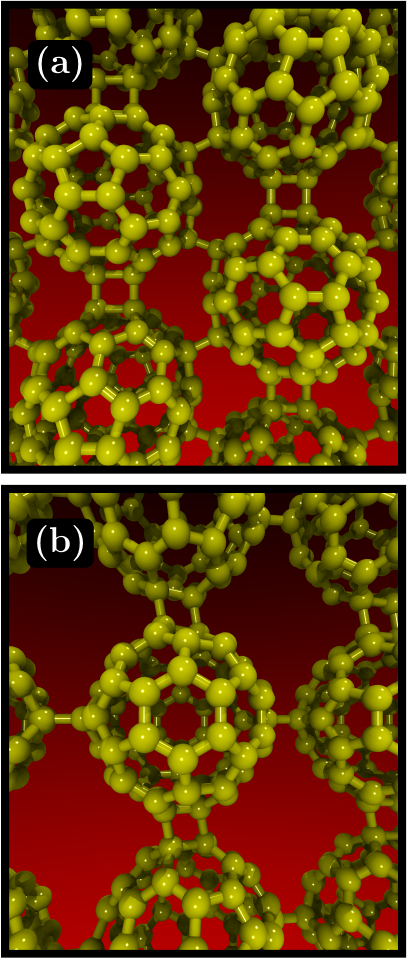}
    \caption{Top view of the atomic structure of the two experimentally obtained Monolayer Fullerene Networks. (a) and (b) display the qHP and qTP phases, respectively.}
    \label{fig:C60}
\end{figure}

2D Fullerene networks are formed by joining fullerene molecules into a monolayer of interconnected units. Depending on the intermolecular connection, different phases can be obtained, such as rhombohedral (R), quasi-hexagonal (qHP), tetragonal (T), or quasi-tetragonal (qTP) ones \cite{xu1995theoretical,li2023superior,hou2022synthesis,kuzmany2004raman}. There are different methods to induce the polymerization of adjacent C$_{60}$ units to create a network \cite{eklund2013fullerene}, using, for example, photons \cite{rao1993photoinduced,yamamoto2007photon}, electron-beam \cite{nakaya2004fabrication}, plasma \cite{takahashi1993plasma,tran2020plasma}, pressure, and pressure-temperature \cite{iwasa1994new,sundar1996pressure,narymbetov2003crystal}. Figure \ref{fig:C60} displays the structure of two experimentally obtained 2D fullerene networks, which are discussed in the present work. Namely, Figures \ref{fig:C60}(a) and \ref{fig:C60}(b) show, respectively, the qHP and the qTP phases of this 2D carbon allotrope.

\subsection{Theoretically Predicted Properties}

From 1993 to 1995, theoretical studies investigated the polymerization of C$_{60}$ molecules, although the formation of dimers, trimers, tetramers, or 3D FCC solids dominated the early research \cite{strout1993theoretical,menon1994structure,adams1994polymerized,pederson1995fullerene,porezag1995structure,okada1999new}. However, the experimental synthesis of a structure composed of stacked planes of polymerized fullerene molecules (similar to graphite) \cite{iwasa1994new,nunez1995polymerized}, stimulated theoretical investigation on a 2D rhombohedral phase of interconnected fullerenes \cite{xu1995theoretical}. Using a carbon tight-binding potential, the authors predicted a semiconductor behavior for the lowest energy configuration of this phase. The structure had energy 2.71 eV/molecule higher than pristine C$_{60}$ and an equilibrium distance of 9.17 {\AA}, in close agreement with the experimental value of 9.22 {\AA} \cite{iwasa1994new}. Subsequently, theoretical predictions on the energetic stability and the electronic structure of R and T phases of C$_{60}$ polymers were conducted by Okada and Saito \cite{okada1999electronic}, using DFT calculations. According to their results, networks in both phases are semiconductors with small indirect gaps. In addition, due to the small interlayer distance, both networks feature a three-dimensional electronic structure, in agreement with the results of Davydov and colleagues \cite{davydov1998packing}.

More recently, after the successful synthesis of qHP-C$_{60}$, several theoretical studies considered various properties and applications of this material. For example, using DFT, Tromer \textit{et al.} \cite{tromer2022dft,junior2022thermal} investigated the mechanical and optoelectronic properties of the qHP-C$_{60}$ monolayer. Their calculations revealed that this fullerene network exhibits an anisotropic elastic modulus and predicted that it could be utilized as a UV collector for photon energies up to 5.5 eV. In the same year, Yu et al. \cite{yu2022anisotropic} also investigated this same structure. Their DFT simulations predicted a direct electronic band gap of 1.55 eV, which is consistent with the experimental results. They also observed that the optical, thermoelectric, and mechanical properties of qHP-C$_{60}$ are anisotropic. In addition, they identified that this anisotropy was due to asymmetry in the bridging of C$_{60}$ units.

Posterior theoretical calculations suggested the use of monolayer qTP- and qHP-C$_{60}$ as photocatalysts for water splitting \cite{peng2022monolayer}. Using semilocal DFT and hybrid functional simulations, this work found that the monolayer could act as either an electronic donor (qHP) or a receptor (qTP) in photocatalysis. Both materials exhibited excellent exciton binding energies, and it was verified that the overall water splitting could occur spontaneously in qTP-C$_{60}$ upon photon excitation at room temperature and acidic conditions. Furthermore, Y. Tong \textit{et al.} \cite{tong2023monolayer} examined the suitability of the tetragonal phase fullerene monolayer for hydrogen separation. The authors found that it exhibited the best pore size match and entropic selectivity, suggesting promise for this material in applications such as H$_2$ purification and water splitting.

Recently, using first-principles calculations, Peng and Pizzochero\cite{Peng2025} tested and confirmed the outstanding, and highly tunable, properties of experimentally observed fullerene structures. In that study, the authors investigated the structural stability of monolayer phases, thermal expansion behavior, and several mechanical properties. Furthermore, that study investigates the photocatalytic properties of such systems, which are experimentally supported. 

The effect of external electric field modulation on the photoelectric properties of the qTP- and qHP-C$_{60}$ monolayers was also investigated \cite{miao2023simulation}. The authors observed the sensitivity of the optoelectronic properties of the monolayers to an applied voltage, finding a transition from a semiconducting to a conducting state for voltages of 25.5 V (qHP) and 30 V (qTP). Their results showed that the visible light absorption peak in a qTP phase was first redshifted and then blueshifted as the electric field intensity increased. Meanwhile, for the qHP phase, the UV peak was blueshifted. Their results show that both materials could potentially be employed in optoelectronic devices, as external electric fields modulate their photoelectric properties.

Later theoretical works also confirmed the superior mechanical \cite{peng2023stability,mortazavi2023structural} and thermoelectric properties \cite{li2023superior} of monolayer fullerene networks. Furthermore, recent studies have significantly advanced the understanding of the properties and possible applications of monolayer C$_{60}$ phases \cite{ajori2024graphullerene,cassiano2024large,capobianco2024electron,champagne2024strongly,shi2024tunable,wang2024strain,suo2024covalent,xu2024photoinduced,alekseev2025thermal}.

\subsection{Synthesized Fullerene Networks}

The first report on polymerized fullerene networks dates back to 1993, when Rao \textit{et al.} reported the creation of a covalently bonded fcc structure from $C_{60}$ molecules through photopolymerization induced by visible or UV light \cite{rao1993photoinduced}. In the next year, Iwasa \textit{et al.} \cite{iwasa1994new} also succeeded in producing 3D fullerene networks, following an alternative process. In 1995, Nuñez \textit{et al.} \cite{nunez1995polymerized} were one of the first to report the synthesis of 2D fullerene phases. They obtained three new C$_{60}$ phases, which included an orthorhombic (O) unidimensional chain (spacegroup \textit{Immm}). The other phases were 2D: one rhombohedral (\textit{Immm}) and one tetragonal (\textit{R$\bar{3}$m}). Synthesis was achieved through a 2 + 2 cycloaddition reaction of double bonds, utilizing moderately high pressures and temperatures. This resulted in a product with a mixture of phases, which was resolved by electron diffraction patterns and X-ray techniques.

In 1998, the tetragonal phase was investigated by Davydov \textit{et al.} \cite{davydov1996pressure,davydov1998tetragonal}. Until then, the T phase was obtained only in mixtures containing O and R phases and was considered metastable. By using different temperature and pressure paths, targeted at 873 K and 2.2 GPa, the authors successfully synthesized a 3D solid composed almost entirely of T-phase layers \cite{davydov1998tetragonal}. In the following years, Chen \textit{et al.} performed the first single-crystal X-ray refinement on the structure of both T \cite{chen2002single} and R \cite{chen2002first} phases. For the T phase, their X-ray structural analysis revealed that the polymer actually crystallized in the orthorhombic space group (\textit{Immm}).
For the R phase, they identified that the polymerization of C$_{60}$ molecules occurred two-dimensionally along the (1 1 1) plane of the fcc packing, suggesting that this process should occur rapidly once it starts at high pressure and temperature.

In 2003, Narymbetov \textit{et al.} \cite{narymbetov2003crystal} obtained single crystals of the 2D polymerized C$_{60}$ T phase without oriented domains under high temperature and pressure. They obtained interlayer distances between T films contrary to previous theoretical predictions \cite{xu1995theoretical,oku2001formation,zhu2002far}, due to partial disorder in the structure, which consisted of 84\% P4$_2$/\textit{mmc} and 16\% \textit{Immm} layers. In 2009, Kazachenko and Ryazanov \cite{kazachenko2009structure} investigated fullerene thin films obtained via the electron-beam dispersion (EBD) method, which combined an electron-induced sublimation of the C$_{60}$ solid (fullerite) with an electron-assisted polymerization. They obtained thin films with different contents of polymerized and almost pristine C$_{60}$s.

By 2010, structures composed of polymerized fullerenes were well-established. However, a number of those were not pure C$_{60}$ polymers, having other species (as alkali metals) or polymers inducing the connection between C$_{60}$ units \cite{murphy1992synthesis,chen2002first,kuzmany2004raman,nakaya2004fabrication,harris2020fullerene}. Furthermore, an all-carbon network had not been isolated as a 2D material or produced in significant quantities \cite{chen2002first,hou2022synthesis}. It was only in 2022 that the synthesis of a pure carbon 2D fullerene network was successful when  Hou \textit{et al.} obtained large single-crystals of monolayer quasi-hexagonal-phase fullerene (qHP-C$_{60}$). In the same article, they also reported the synthesis of a few-layer quasi-tetragonal phase fullerene (qTP-C$_{60}$) \cite{hou2022synthesis}. This study proposed an organic cation slicing procedure to exfoliate the 2D layers from the bulk material, as it cannot be mechanically exfoliated nor cleaved in polar solvents due to the strong Mg-C interaction. Note that the Mg ions were introduced during the synthesis of the bulk crystal, linking adjacent layers together, but were removed through the exfoliation process.

In 2023, evidence for the excellent performance of few-layer fullerene networks as catalysts for water splitting into H$_2$ was reported by Wang and colleagues \cite{wang2023few}. In their work, they synthesized the 2D qHP-C$_{60}$ network following the method by Hou \textit{et al.} and then conducted photocatalytic experiments in pure water. They confirmed that qHP-C$_{60}$ nanosheets display great activity, unlike both pristine C$_{60}$ and bulk Mg$_4$-C$_{60}$. Their results confirmed the theoretical predictions of Peng \cite{peng2022monolayer} and Tong \cite{tong2023monolayer}. In the same year, another breakthrough was reported by Meirzadeh and colleagues \cite{meirzadeh2023few}. They developed a chemical strategy to grow macroscopic thin flakes of the qHP phase that could be mechanically exfoliated. They called these monolayers, which have crystal sizes in the order of hundreds of micrometers, graphullerene.

Another exciting discovery occurred in 2024 when X. Chen \textit{et al.} \cite{chen2024ultrathin} reported the experimental synthesis of a porous fullerene membrane. This was achieved by modifying a qHP sheet, using 600°C heating to promote depolymerization of Mg-intercalated qHP-C$_{60}$. Their process introduced nanopores of size ranging from 1.2 to 5.3 nm, allowing for ultrafast permeation of organic solvents. A similar process was also reported by F. Pan \textit{et al.} \cite{pan2023long}. He and collaborators \cite{he2024nanoporous} reported further advancements in the synthesis of nanoporous fullerene networks, where a centimeter-sized nanoporous 2D carbon monolayer was obtained by rapid pyrolysis of a fullerene monolayer. 

Most recently, in 2025, Zhang \textit{et al.} \cite{zhang2025electrochemical} reported a facile electrochemical process for the gram-scale synthesis of C$_{60}$ monolayers. The process was similar to that employed by Hou \textit{et al.} \cite{hou2022synthesis}, with the addition of a hydrogen-assisted electrochemical exfoliation strategy. The authors reported high-quality and high-yield qHP-C$_{60}$ production. Their scalable process achieved significant production rates of up to 5 g/day, opening the possibility of large-scale exploration of C$_{60}$ polymeric monolayers. The advancements in synthesis methods, coupled with decreasing C$_{60}$ prices (at 20 USD/gram in 2024 \cite{chang2024recent}), generate enthusiasm that is driving the research on 2D fullerene networks. 

\section{Conclusions}
\label{section:conclusion}

The field of 2D carbon allotropes has evolved significantly beyond graphene, encompassing a growing family of experimentally realized networks with diverse structural and electronic characteristics. In this review, we have discussed several noteworthy 2D carbon materials, including monolayer amorphous carbon (MAC), $\gamma$-graphyne (GY), graphdiyne (GDY), biphenylene carbon-based networks (BPN), and 2D fullerene networks. Each of these materials exhibits unique topological motifs, synthesis routes, and physical properties that are expanding the design space for carbon-based nanotechnologies.

The emergence of MAC has demonstrated that even disordered atomic networks can preserve key features, such as high conductivity and mechanical resilience, while introducing tunable electronic band gaps, a feature absent in pristine graphene. Likewise, GYs and GDYs offer direction-dependent properties and intrinsic porosity, enabling promising applications in electronics and molecular sieving. BPC structures are mechanically robust, while presenting regularly spaced pores that have been considered promising for gas separation and water desalination, for instance. Also, their electronic structure exhibits a range of conducting behaviors, transitioning from metallic to semiconducting. Additionally, some BPC-based materials have been identified as high-order topological insulators, opening up a new set of possible technological applications. On the other hand, quasi-periodic C$_{60}$ fullerene networks expanded the concept of 2D carbon into curved topologies and molecular packing.

Despite substantial advances, challenges remain. Large-scale, defect-free synthesis remains challenging for many of these allotropes, and their long-term stability under operational conditions remains unclear. Moreover, the theoretical understanding of disorder, interfacial phenomena, and many-body interactions in these novel systems is still in progress.

Overall, the growing landscape of 2D carbon allotropes presents a rich platform for fundamental research and emerging technologies, marking a golden new era in carbon-based materials science.

\section*{Acknowledgements}
This work received partial support from the Brazilian Coordination for the Improvement of Higher Education Personnel (CAPES), the National Council for Scientific and Technological Development (CNPq), Research Support Foundation of the Federal District (FAPDF), and Research Support Foundation of São Paulo (FAPESP).  
R.P. acknowledges Brazilian funding agencies: FAPESP (grant \#2021/14977-2) and CNPq (grant \#313592/2023-3).
A.F.F. is a fellow of the Brazilian Agency CNPq-Brazil (\#302009/2025-6) and acknowledges grant \#2024/14403-4 from S\~ao Paulo Research Foundation (FAPESP). MB and LDM acknowledge the support of the Brazilian Research Agencies CAPES and CNPq.
M.B. and LDM acknowledge the support of the Brazilian Research Agencies CAPES and CNPq (grant \#313065/2023-3).
G.S.L.F. thanks the postdoc scholarship financed by the São Paulo Research Foundation (FAPESP) (grant \#2024/03413-9)
M.L.P.J. acknowledges financial support from FAPDF (grant 00193-00001807/2023-16), CNPq (grants 444921/2024-9 and 308222/2025-3), and CAPES (grant 88887.005164/2024-00). Computational resources were provided by the National High-Performance Computing Center in São Paulo (CENAPAD-SP, UNICAMP, projects proj931 and proj960) and the High-Performance Computing Center (NACAD, Lobo Carneiro Supercomputer, UFRJ, project a22002).
L.A.R.J. acknowledges the financial support from FAP-DF grants 00193.00001808/2022-71 and 00193-00001857/2023-95, FAPDF-PRONEM grant 00193.00001247/2021-20, CNPq grants 350176/2022-1, 167745/2023-9, 444111/2024-7, and PDPG-FAPDF-CAPES Centro-Oeste 00193-00000867/2024-94.
We thank the Coaraci Supercomputer Center for computer time (FAPESP grant \#2019/17874-0) and the Center for Computing in Engineering and Sciences at Unicamp (FAPESP grant \#2013/08293-7).

\printcredits
\bibliographystyle{unsrt}
\bibliography{cas-refs}

\end{document}